\documentclass[aps,prb,floatfix,superscriptaddress,showpacs,11pt]{revtex4}
\usepackage[T1]{fontenc}
\usepackage[utf8]{inputenc}
\usepackage{amsmath}   
\usepackage{amssymb}
\usepackage[amsmath,amsthm,thmmarks,hyperref]{ntheorem}
\usepackage{etex}
\usepackage{multirow}
\usepackage{verbatim}
\usepackage{graphicx}   
\usepackage{subfigure} 
\usepackage{tikz}
\usetikzlibrary{arrows}
\usetikzlibrary{patterns}
\usepackage{xcolor}
\usepackage[unicode,colorlinks,plainpages=false]{hyperref}
\usepackage{makeidx}
\usepackage{ae}

\usepackage{bookmark}

\begin{document}
\title{Micron-scale deformation: a coupled in-situ study of strain bursts and acoustic 
emission}
\author{Ádám István Hegyi}
\email{hegyi@metal.elte.hu}
\affiliation{Department of Materials Physics, Eötvös University, Pázmány Péter sétány 1/a,
H-1117 Budapest, Hungary}
\author{Péter Dusán Ispánovity}
\affiliation{Department of Materials Physics, Eötvös University, Pázmány Péter sétány 1/a,
H-1117 Budapest, Hungary}
\author{Michal Knapek}
\affiliation{Charles University in Prague, Faculty of Mathematics and Physics
Department of Physics of Materials, Ke Karlovu 5 121 16 Prague 2}
\author{Dániel Tüzes}
\affiliation{Department of Materials Physics, Eötvös University, Pázmány Péter sétány 1/a,
H-1117 Budapest, Hungary}
\author{Krisztián Máthis}
\affiliation{Charles University in Prague, Faculty of Mathematics and Physics
Department of Physics of Materials, Ke Karlovu 5 121 16 Prague 2}
\author{Franti\v{s}ek Chmelík}
\affiliation{Charles University in Prague, Faculty of Mathematics and Physics
Department of Physics of Materials, Ke Karlovu 5 121 16 Prague 2}
\author{Zoltán Dankházi}
\affiliation{Department of Materials Physics, Eötvös University, Pázmány Péter sétány 1/a,
H-1117 Budapest, Hungary}
\author{Gábor Varga}
\affiliation{Department of Materials Physics, Eötvös University, Pázmány Péter sétány 1/a,
H-1117 Budapest, Hungary}
\author{István Groma}
\affiliation{Department of Materials Physics, Eötvös University, Pázmány Péter sétány 1/a,
H-1117 Budapest, Hungary}

\date{\today}

\begin{abstract}
Plastic deformation of micron-scale crystalline materials differ considerably from 
bulk ones, because it is characterized by random strain bursts. To obtain a detailed picture about 
this stochastic phenomenon, micron sized pillars have been fabricated and compressed in the 
chamber of a SEM. An improved FIB fabrication method is proposed to get non-tapered
micro-pillars with a maximum control over their shape. The in-situ compression device used allows high accuracy sample positioning and force/displacement measurements with high data sampling rate. The collective avalanche-like motion of dislocations appears as stress 
drops on the stress-strain curve. To confirm that these stress drops are directly related to  
dislocation activity, and not to some other effect, an acoustic emission transducer has been 
mounted under the sample to record emitted acoustic activity during strain-controlled compression tests of Al-5\% Mg micro-pillars. The correlation between the stress drops and the 
acoustic 
emission signals indicates that indeed dislocation avalanches are responsible for the stochastic 
character of the deformation process.
\end{abstract}


\maketitle

\section{Introduction}
During the past decades miniaturizing mechanical and electronic devices inspired 
research on determining the mechanical properties of micron sized specimens 
\cite{c3,c4,c5,c6,c7}. The vast majority of micro-electromechanical sensors contain micrometer size 
pieces, for instance, accelerometer measures the buckling of a micro-cantilever or chemical 
and biological sensors use cantilever transducers as a platform  \cite{c1,c2}. In order to be 
able to design devices that are even smaller, the detailed physical properties of the deformation 
processes should be studied. 

The plastic deformation of crystalline materials is usually caused by the motion of the dislocations. 
In macroscopic samples, due to the large number of dislocations, the stress-strain response of the 
material is smooth allowing to predict the properties of the material with a high accuracy.  In 
contrast, at the micrometer scale the inhomogeneity of the dislocation microstructure is in 
the order of the sample size  resulting non-deterministic response due to the stochastic motion of 
dislocations \cite{c8,c9,c10,c11,c12}. So, in order to design new microscopic devices the 
mechanical properties of materials have to be described by a statistical 
approach at this scale. 

The first evidence of intermittent crystal plasticity was obtained  on ice single crystals by 
detecting  strong acoustic emission signals during creep deformation\cite{c13,c14}. A decade ago 
by compressing pure single crystalline Ni micro-pillars Dimiduk et.~al.~found that instabilities in the form of strain jumps dominate micrometer scale crystal plasticity\cite{c15, c16, c17} 
raising 
the questions (i) what is the limit between microscopic and macroscopic deformation, and (ii) how one can 
define material strength parameters, like flow stress, for micron-scale objects  
\cite{c18,c19,c28}.

Due to its statistical nature to determine the properties of micro-deformation a vast amount of
micrometer size samples are needed. One of the easiest and most frequently applied method to fabricate them is focused ion beam (FIB) milling 
\cite{c20}, with
the main advantage that one can meanwhile visually check the process of ion milling. To 
shorten the fabrication time of the micro-pillars several different growing (FIB-less) methods were 
developed \cite{c21, c22}. These, however, do not allow to produce pillars from any type of 
material, and to have control over the initial dislocation content in the sample. Moreover, for pillars 
produced by growing the connecting force between the substrate and the micro-pillar itself can be 
rather weak.\cite{c23}

To fabricate micro-pillars by FIB milling two approaches are commonly applied: 
``lathe'' and ``top-down'' millings \cite{c24}. Lathe-milling uses ion beam (almost) perpendicular to the 
axis of the micro-pillar’, and the pillar is rotated around to get the cylindrical shape. Usually 
this procedure  is used on the side of a thin layer of a crystal material. In the top-down 
technique the pillar axis is parallel to the ion beam, the ions etch the surface of the sample 
prepared. In this case, however, the height of the micro-pillar and its tapering are poorly 
controlled. The new method outlined below    combines the advantages of both fabrication techniques. 
 With this procedure a non-tapered micro-pillar can be obtained anywhere on the surface of a bulk 
material, and much less preparation time is required \cite{c27}. Moreover, by this technique it is 
possible to investigate in-used parts in a quasi non-destructive way. 

In the studies presented our main goal was to get an insight into the fundamentals of plastic 
deformation at the micrometer scale. The new fast micro-pillar fabrication procedure developed gives us 
the chance to perform  a statistical analysis of the deformation processes. This is essential 
because 
during an individual compression experiment the stress varies intermittently (see below), meaning 
that the dislocation system gives stochastic response to the acting force \cite{c11, c14, c16, c25, c26}. 
As a result of this defining material parameters from a unique measurement is impossible. So, 
hardness, or strain hardening can only be calculated from a statistical analysis \cite{c26, c28}.

The compression tests were performed on Al-5\% Mg alloy micro-pillars fabricated onto 
the surface of the bulk material. An important feature of this alloy is that it shows the Portevin-Le 
Chatelier (PLC) effect \cite{c31,c32,c33,c34}.  So, even for bulk samples, intermittent 
stress-strain response is observed which is related to the pile-up and break out of the 
dislocations from the solute atom atmosphere acting as junctions for mobile dislocations \cite{c33, 
c34}. This mechanism also generates strong acoustic signals with energies slightly higher 
than those caused by dislocation avalanches. Since the stress drops caused by PLC and dislocation 
avalanches are clearly distinguishable, the tests on this material allowed us to determine the  
sensitivity of our acoustic emission detector setup.  

\section{Sample preparation}

Prior to fabrication the size of the micro-pillar has to be decided. Since in this 
study we are interested in the collective dislocation phenomena, we have to have enough dislocations 
in the system initially. On the same time the sample size should not be too large to hinder the 
stress drops. Normally in fcc metals like Al the dislocation density varies between  
$10^{11}-10^{14}$ $\text{m}^{-2}$, therefore, the average spacing between dislocations is about $3-0.1$ $\mu$ 
m. 
Since the dislocations tend to form a cell like structure with a characteristic size of about 10 
times the dislocation spacing, the pillar size is selected to be in the order of the cell size.

In-situ micro-pillar deformation tests demand very careful and precise sample preparation. To get 
the required surface properties, orientation, and initial dislocation density the following steps 
were performed: after a short etching the Al-5\% Mg sample was electropolished in 
perchloric electrolyte $D2$ solution with 60 mA/mm$^{2}$ current density. The lattice 
orientation was 
measured using electron back-scattered diffraction (EBSD). The sample was cut by electric 
discharge machining (EDM) to have a surface orientation $\left\langle 1,2,3 \right\rangle $.  
After this, another electropolishing  step was made, then the sample was heat treated for $72$ 
hours at $200$ $^\circ$C. After the heat treatment, the surface was electropolished again with 
30 mA/mm$^{2}$ and the orientation was checked again by EBSD. The sample was pre-deformed 
parallel to the $\left\langle1,2,3 \right\rangle $ axis to 20 MPa. The initial  
dislocation density was measured by TEM and X-ray line profile analysis, and was found to be 
$2 \times 10^{13}$ m$^{-2}$. Taking this value into account, a pillar geometry with rectangular cross 
section of $4\times 4$ $\mu$m$^2$ was selected with a height of 12 $\mu$m corresponding to a aspect 
ratio about 3:1:1 commonly applied in earlier studies.

\begin{figure}[!ht]
\begin{center}
\includegraphics[scale=0.2]{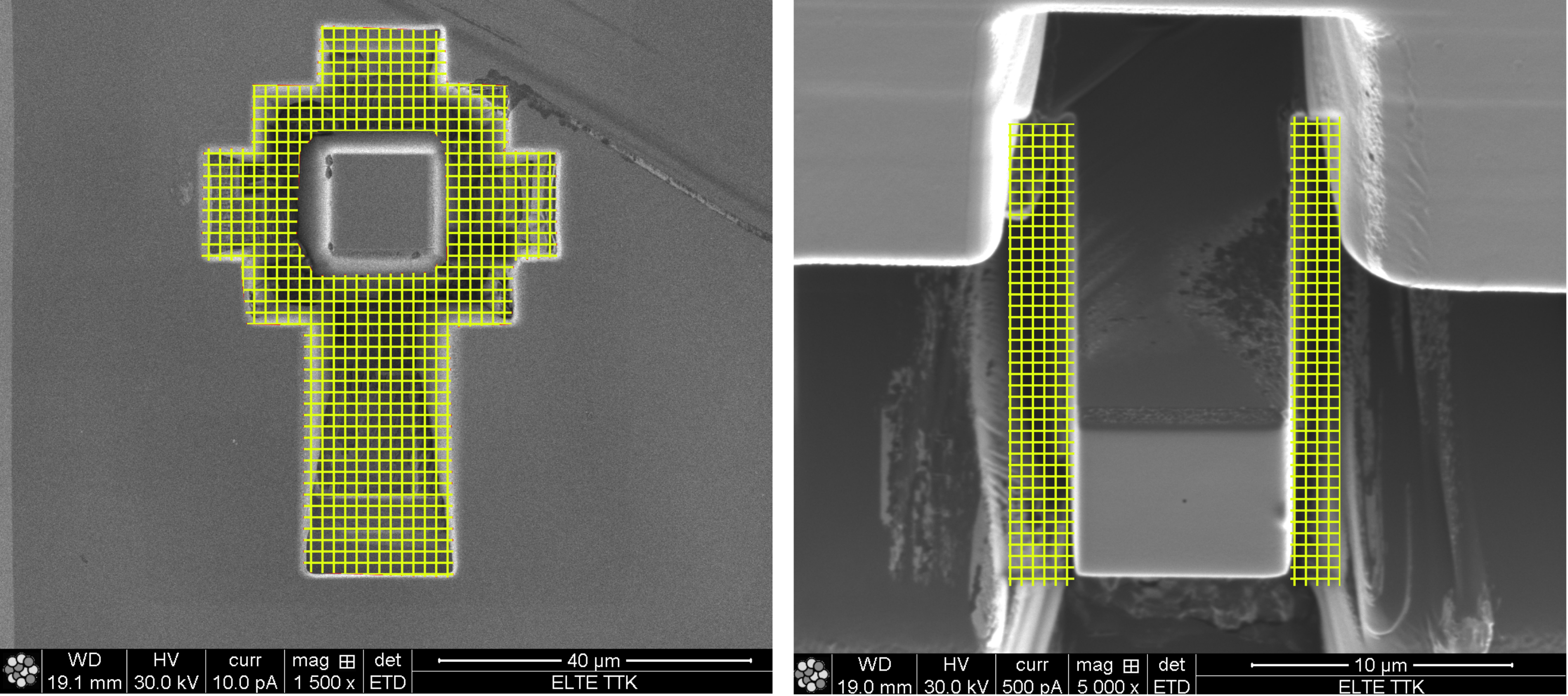}
\begin{picture}(0,0)
 \put(-310,126){\color{white} \textsf{(a)}}
\end{picture}
\begin{picture}(0,0)
 \put(-155,126){\color{white} \textsf{(b)}}
\end{picture}
\it\caption{\label{fig1} a) The initial FIB milling step marked by the grid to fabricate the 
``raw'' pillar and the hole around it, that is necessary to carry out the compression test. At this  milling 
step $30$ nA
 ion current is applied. b) The second finalizing milling step with $45^{\circ}$ ion 
direction angle (see Fig.~2, too). The area removed by this step is marked by the grid. The ion 
current is $5 nA$. Both pictures are taken from the ion direction.}
\end{center}
\end{figure}
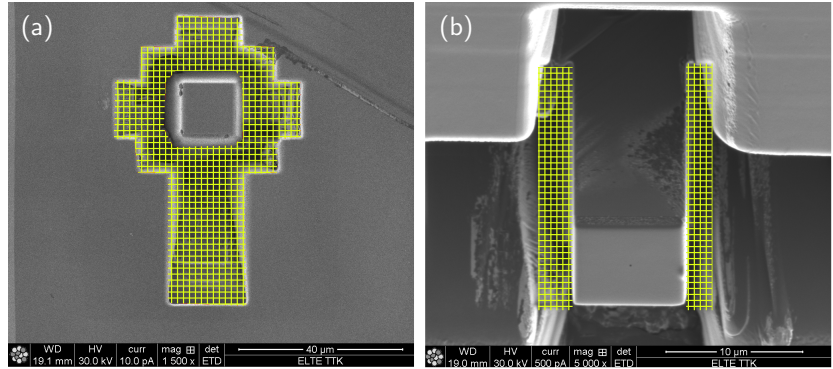

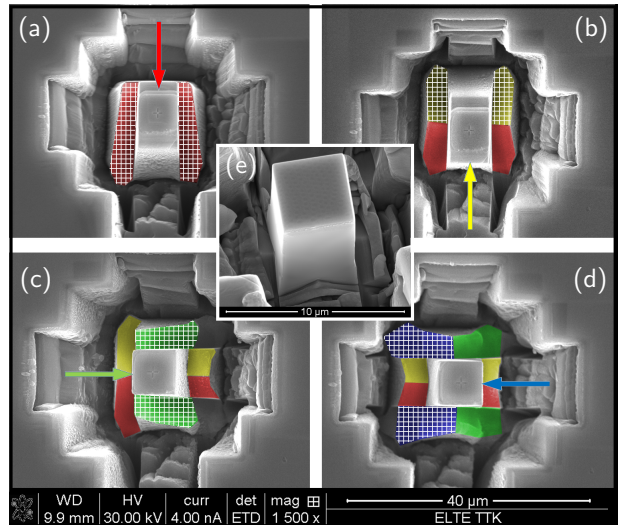
\begin{figure}[!ht]
\begin{center}
\includegraphics[scale=0.2]{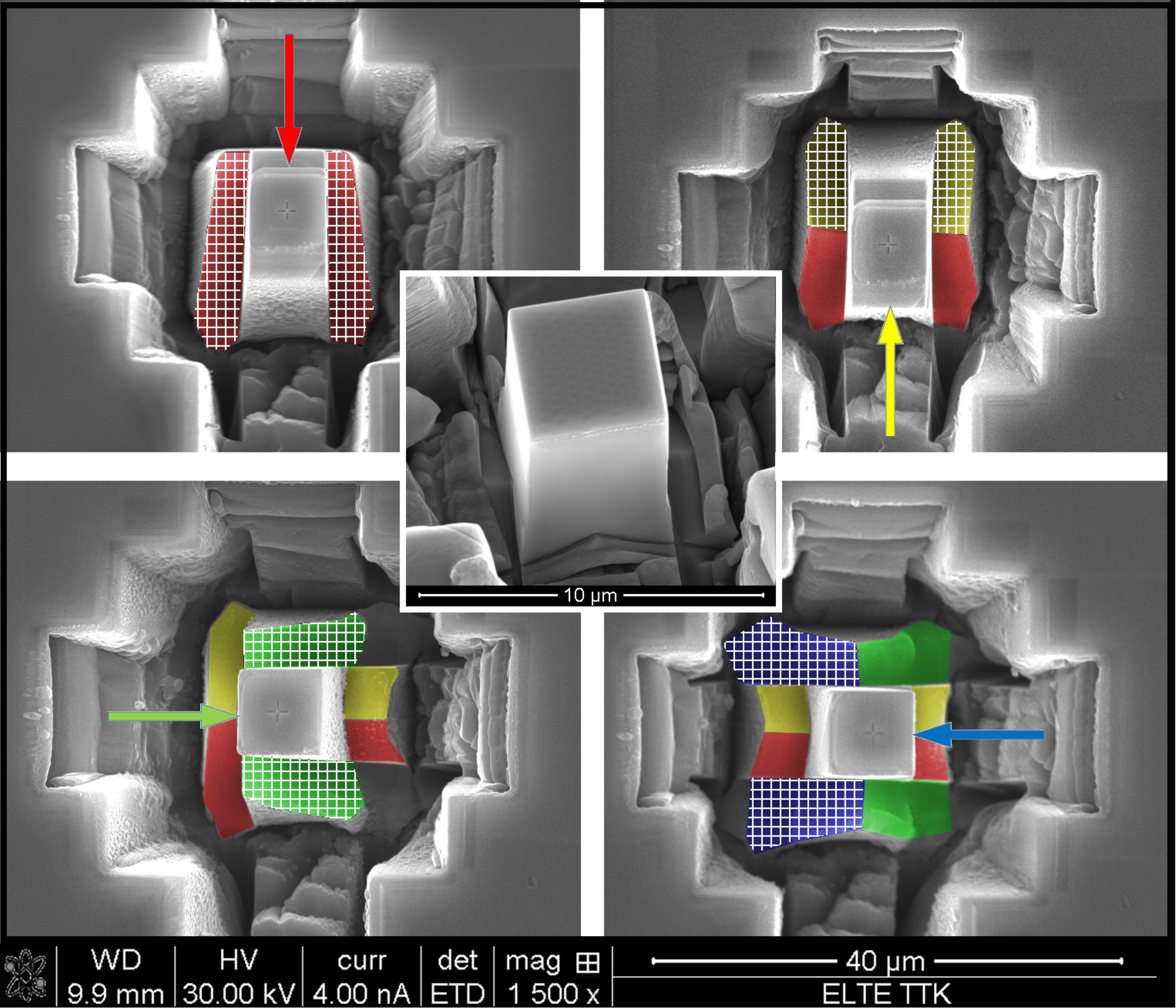}
\begin{picture}(0,0)
 \put(-230,185){\color{white} \textsf{(a)}}
\end{picture}
\begin{picture}(0,0)
 \put(-23,185){\color{white} \textsf{(b)}}
\end{picture}
\begin{picture}(0,0)
 \put(-236,90){\color{white} \textsf{(c)}}
\end{picture}
\begin{picture}(0,0)
 \put(-30,90){\color{white} \textsf{(d)}}
\end{picture}
\begin{picture}(0,0)
 \put(-165,133){\color{white} \textsf{(e)}}
\end{picture}
\it\caption{\label{fig2} The details of the second finalizing step of the micro-pillar fabrication 
procedure with an ion current of 5 nA. a) After the first step seen in Fig.~1 the sample is tilted by 
$45 ^{\circ}$ resulting a tilted ion beam direction. After this, two rectangular FIB patterns 
(marked by the red grid) are used to get the surface marked by red. b) As a next step 
the sample is rotated by $180^{\circ}$, and two rectangular FIB patterns are used again to get the 
surface marked by yellow. With a rotation by $90^{\circ}$, and repeating the two previous steps   
the surfaces marked by green c) and blue d) are obtained. With these a square shaped pillar is 
achieved. In order to get a pillar with smooth surface and practically no tapering the process is repeated 
with ion currents  1 nA, and 100 pA too. The final pillar obtained is seen in the inset e).}
\end{center}
\end{figure}

After the above mentioned sample preparation processes a ``surrounding hole'' was milled by 30 
nA ion current  around the pillar. The FIB milling pattern used is 
marked by the grid in Fig.~\ref{fig1}/a. The sample was oriented so that the normal vector of the 
surface was parallel to the ion beam direction. Then a thin Pt layer was deposited onto the top surface 
of the micro-pillar. The cap helps the ion beam to fabricate smooth side surface of the 
pillar and due to its  amorphous structure it is very hard which can help to eliminate 
effects related to the missalignment of the compressing tip (see below). To do the next step the 
stage is tilted by $7^{\circ}$.  Due to the $52^{\circ}$ ion-electron beam angle this 
result in a $45^{\circ}$ ion beam direction compared to the surface normal. To avoid flaws originating from the non-repeatability of the microscope stage during positioning a 2 $\mu$m size cross was drawn 
to the geometric center of the top surface of the micro-pillar.  After this the milling 
steps explained in the caption of Fig.~\ref{fig2} were performed. In order to further decrease 
tapering 
\cite{c36}, a final ``polishing like'' step is performed  with 30 pA ion beam over-tilted by 
$1^{\circ}$ compared to the pillar axis.

Apart from the first step explained above the fabrication works with high surface angle. This 
increases the effectiveness of the milling by a factor of  2.5-3 compared to the perpendicular beam 
setup commonly used  \cite{c35}. The $45 ^{\circ}$ milling direction applied in the second phase 
lets us to fabricate micro-pillars anywhere on the flat sample surface. 

It has to be mentioned, that the Ga ion beam forms an irradiated skin on the micro-pillar. The 
slightly tilted final ``polishing'' step with the reduced 10 pA ion current and the lowered acceleration voltage  (to 20 kV 
or 10 kV)  decreases the thickness of this layer leading to
a negligible effect at pillar size used \cite{c38}.

To sum up the most important advantages  of the milling procedure proposed are: (i) pillars can be 
fabricated at any position on the surface, (ii) the preparation is touchless, so, the damage or the 
deformation of the pillar can be avoided during the whole production process, and (iii) effects related to tapering and Ga implantation can be neglected.


\section{In-situ device}

For an in situ micromechanical test the testing device has to fit into the vacuum chamber
of the SEM. Such compression devices are commercially available, yet we used an
inexpensive solution. The device easily fits into the vacuum chamber of a FEI Quanta
3D SEM.

Two linear ultrasonic motors are used for the $X$ and $Y$ positioning of the sample. The AE transducer is mounted on the top of the two stages. In $Z$ direction  two 
stages are used. One is a linear step-motor stage used for the ``raw'' motion of the compressing 
tip to get it close to the sample. The other one mounted on the linear step-motor stage is
a piezoelectric positioning (PEP) stage with about 0.1 nm resolution. During the actual 
compression test only this stage is moved.  To measure the external force a standard spring mounted 
on the PEP stage is used with strong transverse but very weak longitudinal stiffness. The 
elongation $E$ of the spring is measured by a capacitive sensor with 0.1 nm resolution. If the 
PEP stage is moved by the distance $P$, and the capacitive sensor measures $E$ elongation than the sample deformation is  $D = P - E$, and the acting force is $F=SE$, where $S$ is the stiffness of the spring. The pillar 
compression 
is performed using a flat punch diamond tip. It should be mentioned that to avoid the charging of the 
compressing head in the SEM a tip doped by boron has to be used.

For being able to measure instabilities related to PLC effect and dislocation avalanches it is 
crucial to have a fast enough feedback controlling system and minimum of $1kHz$ data collection 
rate. These are achieved by an analogous PID type feedback electronics and a fast 16bit AD 
converter. 

The range and the resolution parameters of the device are summarized in Table 1.

\begin{table}[!ht] \centering \caption{Parameters of the nanodeformation device} \label{my-label} 
\begin{tabular}{l|l|l|l}
Part name & Total range & Resolution & Accuracy \\
\hline X and Y & $\pm 8$ mm & 
0.5 $\mu$m & 0.01 $\mu$m \\
\hline Z coarse & 9 mm & 2 $\mu$m & 0.5 $\mu$m \\
\hline Z fine & 35 $\mu$m & 1 nm & 0.1 nm \\
\hline Force sensor & 20/50 mN & 1/2.5 $\mu$N & 1/2.5 $\mu$N 
\end{tabular} 
\end{table}

To achieve these resolutions the thermal and elastic elongation of the different parts have to be 
negligible during the typically few minutes of the measuring time. For this reason, 
several different additional parts are added to the compression device. One main problem is to 
reduce the heat produced by the motors of the stages. The Quanta 3D has an environmental stage with 
a Peltier sample holder to set the sample temperature. The cold point of this Peltier stage is 
mounted to the bottom of the device. By this, the temperature is stabilized at 15 $^{\circ}$C.  
Another important issue that needs to eliminated is the vibration of the spring of the force sensor arising due to the lack of air, that is, damping. To avoid this rather disturbing effect strong permanent magnets are placed close to the lamellae of the force measuring spring providing the necessary damping due to the eddy currents.

\section{Acoustic emission measurement}

An Acoustic Emission (AE) measuring system was also employed to study the dynamic processes  
occurring during the plastic deformation of micropillars. Acoustic emissions are transient elastic 
waves generated by the rapid release of energy from localized sources within the material. Thus, 
AE signals are generated where sudden localized structural changes take place, like dislocation motion or 
twinning. So, it provides information about the dynamic phenomena involved in plastic deformation 
\cite{c39}. 

In bulk materials, a direct correlation of AE parameters with the stress-strain curves can reveal 
the activation of different deformation mechanisms \cite{c40, c41, c42, kovacs2014statistical}. It is important 
to note that the collective motion of at least several tens of dislocations is 
necessary in order to obtain a detectable AE signal \cite{c43}. So, in terms of AE, a motion of a 
single dislocation is typically ``silent'' and, in turn, a detectable AE signal (if caused by dislocation activity) reflects 
the cooperative character of dislocation motion. 

Crackling or avalanche-like type of plasticity is not only characteristic for micron-scale objects, but also for bulk samples\cite{c42}. In this case the AE technique exhibits a great potential to provide 
information on these dynamic processes invisible on the deformation curves. To the knowledge of the authors, the study presented in the 
paper is the first attempt to record AE signals during the micro-pillar compression. 

The AE signal measurement was performed by a Physical Acoustics PCI-2 
acquisition board based on a continuous storage of AE signals with 2 MHz sampling rate. The 
full scale of the A/D converter was $\pm 10$ V. The AE was amplified by 60 dB in the frequency 
range of $100-1200$ kHz. The background noise during the tests did not exceed 24 dB. Due to 
this small noise level the detecting threshold level was set to 26 dB. The AE was recorded 
simultaneously during uni-axial compression of the micro-pillar. 

A rectangular piece of material (with pillar samples fabricated onto its surface) was attached 
directly to the AE transducer using a metallic spring. In addition, the acoustic contact was 
improved by means of a vacuum grease.

\begin{figure}[!ht]
\begin{center}
\includegraphics[scale=0.3]{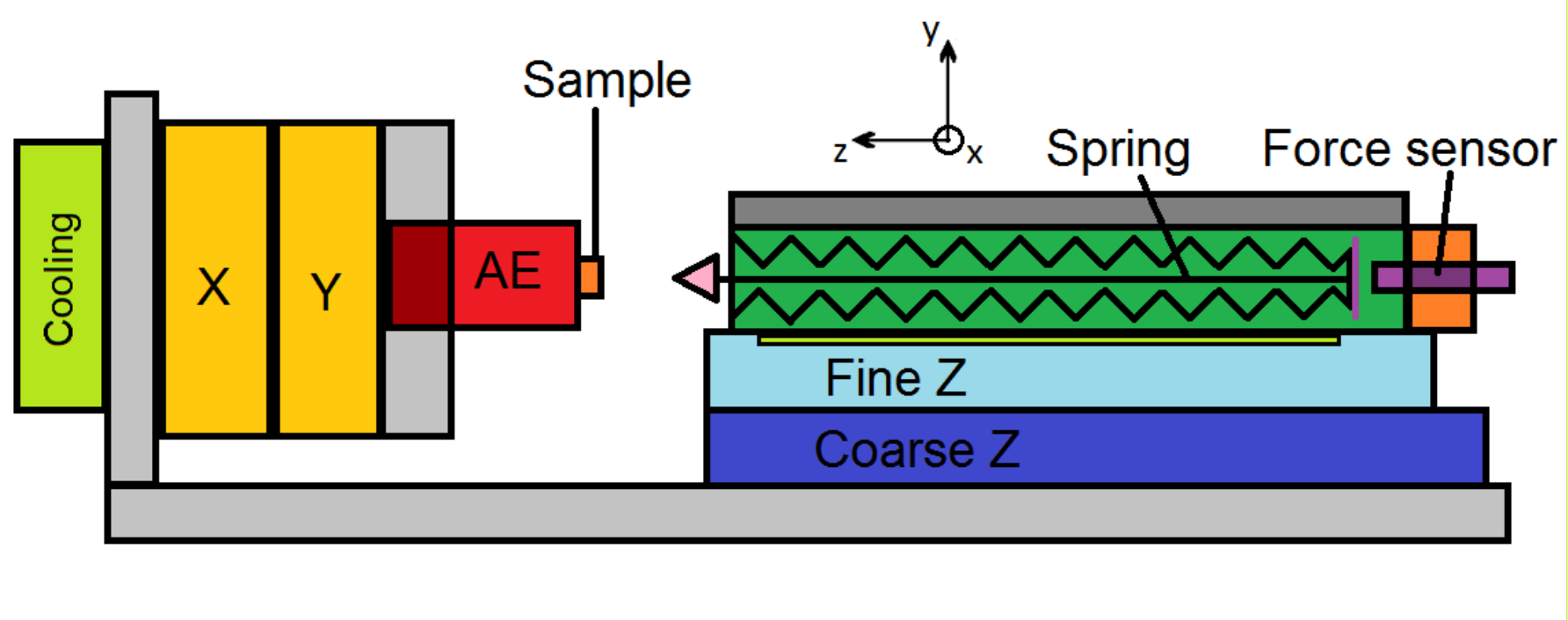}
\it\caption{\label{fig3} Load versus time curve obtained by a micro-pillar compression with 
constant strain rate. In inset a) a stress drop is enlarged. Inset b) shows the waveform of an 
acoustic signal.}
\end{center}
\end{figure}

The load as a function of the time together with the acoustic emission signal obtained on a 
Al-5\% Mg micro-pillar at a constant compressive strain rate is plotted in Fig.~\ref{fig3}.  
As it is expected the sample shows the well known PLC effect. The stress drops at the very beginning 
of the deformation and just before plastic yielding (enlarged in inset a)) can be attributed to the 
breakout of dislocations from the surrounding solute atoms. As it is seen in inset a) just at the 
onset of the stress drop a large acoustic emission signal is detected.     
 
At the micro-pillar size (4x4x12 $\mu m^{3}$) used in the test the PLC effect competes with the 
intermittent dislocation motion (dislocation 
avalanches). This dual effect destroys the well-known periodic stress drop structure of the 
deformation curve, and randomly distributed avalanches can be found. In inset b) the waveform of 
the acoustic signal can be seen. The large peaks on the acoustic emission signal are generated by 
collective motion of many dislocations. 

\section{Summary}

In order to understand in detail the deformation properties of micron sized objects 
experiments carried out on a large ensemble of specimens are needed. The pillar fabrication method presented in the paper is considerably faster
than those applied earlier. This opens the possibility to carry out investigations that 
can reveal the statistical properties of micron scale plasticity. The results obtained indicate
that detecting acoustic emission signal related to the cooperative motion of dislocations is 
feasible even from the small (about 100 $\mu$m$^2$) volume of a micro-pillar.

\begin{acknowledgments} 
Financial supports of the Hungarian Scientific Research Fund (OTKA) under
contract numbers K-105335  and PD-105256 and of the European Commission under grant agreement No. 
CIG-321842 are also acknowledged. PDI is supported by the J\'anos Bolyai Scholarship of the 
Hungarian Academy of Sciences.
\end{acknowledgments}

\bibliography{references}
\end{document}